\documentclass[prb,twocolumn,superscriptaddress,showpacs,%
preprintnumbers,amsmath,amssymb]{revtex4}
\usepackage{amsmath}
\usepackage{graphicx}
\usepackage{dcolumn}

\begin{document}

\title{Grading effects in semiconductor nanowires with longitudinal heterostructures}
\author{Andrey Chaves}\email{andrey@fisica.ufc.br}
\author{J. A. K. Freire}\email{king@fisica.ufc.br}
\author{G. A. Farias}\email{gil@fisica.ufc.br}
\affiliation{Departamento de F\'isica, Universidade Federal do
Cear\'a, Caixa Postal 6030, Campus do Pici, 60455-900 Fortaleza,
Cear\'a, Brazil}
\date{ \today }

\begin{abstract}
The role of graded interfaces between materials in a cylindrical
free-standing quantum wire with longitudinal heterostructures is
theoretically investigated, by solving the Schr\"odinger equation
within the effective mass approximation. Previous work on such
wires with abrupt interfaces have predicted that, as the wire
radius is reduced, the effective potential along the growth
direction is altered and might lead to a carrier confinement at
the barriers, as in a type-II system. Our results show that when
graded interfaces are considered, such potential acquires a
peculiar form, which presents cusps at the interfacial regions,
yielding to electron confinement at interfaces. Numerical results
also show that, in some special cases, interfacial confinement and
type-I to type-II transitions can also be induced by applying a
magnetic field parallel to the wire axis.

\end{abstract}


\maketitle

\section{Introduction}

Semiconductor nanowires have attracted much interest due to their
novel electronic and optical properties owing to their
one-dimensionality and possible quantum confinement effects in two
dimensions. \cite{review} In the past few years, many research
groups have reported the growth of semiconductor nanowires with
longitudinal heterostructures, namely, superlattice nanowires
(SLNW). \cite{Deppert, Solanki} Much theoretical and experimental
study has been made about such wires, where potential applications
of these structures as nano bar-codes, waveguides, light-emitting
diodes and lasers have been suggested. \cite{Gudiksen, Chaves, Li}

Y.-M. Lin {\it et al.} \cite{Dresselhaus} have theoretically
investigated SLNW composed by PbS, PbSe and PbTe, and it was shown
that they are promising systems for thermoelectric applications. A
hybrid {\it pulse laser ablation/chemical vapor deposition}
(PLA-CVD) process was developed by Y. Wu {\it et al.} \cite{Wu}
for the synthesis of single-crystalline nanowires with periodic
longitudinal Si/SiGe heterostructures, and it was suggested that
these nanowires could be used as important building blocks for
constructing nanoscale electronic circuits and devices. The
fabrication of high quality InP/InAs SLNW by chemical beam epitaxy
has been reported by L. Samuelson's research group, \cite{Fuhrer,
Deppert} where it reached a high degree of control of size and
electron number of such systems. A recent publication by Gudiksen
{\it et al.}\cite{Gudiksen} has presented high-resolution
transmission electron microscopy (TEM) images and composition
analysis of a GaAs/GaP SLNW, synthesized by laser-assisted
catalytic growth. It was revealed that the transition between GaAs
and GaP layers is not atomically abrupt, but rather exhibits a
graded interface of 15-20 nm for a $\sim$ 20 nm diameter Au
catalyst. Their results indicate that for a smaller wire radius,
the interface would be reduced, and a 5 nm diameter SLNW, {\it e.
g.}, should have variations of $<$ 5 nm across the junction
interfaces.

Calculations based on the one-band effective mass theory were made
by L. C. L. Yan Voon {\it et al.}, \cite{Voon2} for studying the
electronic states in free standing ({\it i. e.}, not embedded in a
matrix) GaAs/AlAs and InAs/InP SLNW. They predicted the existence
of a barrier localization of longitudinal states, as in a type-II
confinement potential, \cite{Weiser} which is induced by the
strengthening of the radial confinement for thin wires. Their
results indicated the possibility of this modulated structure to
display free-carrier like behavior along the nanowire axis, when a
critical wire radius is considered. These predictions were
confirmed later by M. Willatzen {\it et al.}, \cite{Voon3} and it
has been shown that the existence of critical radii for inversion
of states localization is much more general, and it is also
present in multi-band based calculations. \cite{Voon1} Although
some of the experimental work in the literature has demonstrated
gradual transitions between the heterostructure materials,
\cite{Gudiksen} all of these theoretical investigations on SLNW
have dealt only with models for abrupt interfaces, and the
existence of graded interfaces was neglected.

In the present work, the effective mass approximation is used to
describe the confinement of electrons in a cylindrical
free-standing quantum wire (QWR) with longitudinal
heterostructures, under an applied magnetic field parallel to the
wire axis. In our model, the existence of graded interfaces
between materials is taken into account, and the effective mass is
assumed to depend on spatial coordinates. \cite{king} This model
has been applied to the description of confined states of GaP/GaAs
and InP/InAs QWR. Numerical results confirm the predictions of
previous studies in the literature: indeed, for abrupt interfaces,
as the wire radius becomes thinner, a change in longitudinal
localization of carriers is induced by the creation of a step-like
effective potential, composed by the heterostructure bands
mismatch and the radial confinement energy. In this case, a
critical radius can also be found where the longitudinal effective
potential has a band offset equal to zero, which makes the
electron behave like a free-carrier along the wire axis, despite
the presence of a heterostructure. However, considering smooth
interfaces, the effective potentials along the wire axis acquire a
peculiar form, in which electron states may be confined inside
traps formed at the interfacial regions, even for a small
interface thickness (less than 5 nm). This result, which was not
observed in previous works, allows us to discard any possibility
of having free carrier like behavior in such wires with non-abrupt
interfaces. Our model also shows that, for excited states, such
features of abrupt and non-abrupt QWR can be observed not only by
reducing the wire radius, but also by increasing the intensity of
a magnetic field parallel to the wire axis. This can be an
interesting result for device applications, since it demonstrates
that a change in longitudinal potential and carrier localization
in heterostructured free-standing QWR can be obtained just by
tuning an external parameter, namely, the magnetic field
intensity.

The paper is organized as follows: in Sec. II, we present our
theoretical model for the description of electronic states in
heterostructured QWR with graded interfaces under applied magnetic
fields. In Sec. III, the results for GaP/GaAs and InP/InAs QWR are
presented and discussed. Finally, in Sec. IV, we present our
conclusions.

\section{Theoretical Model}

Our system consists of a circular cylindrical quantum wire, at an
infinite potential region, with a single longitudinal
heterostructure. In cylindrical coordinates, the inclusion of a
magnetic field potential into the Hamiltonian, for a
$\overrightarrow{B}=B\widehat{z}$ field, is made through the
symmetric gauge vector potential, namely
$\overrightarrow{A}=\frac{1}{2}B \rho \widehat{\theta}$. Hence,
the Schr\"odinger equation for this system, within the effective
mass approximation, is given by
\begin{eqnarray}
\Big\{ - \frac{\hbar^2}{2 m^{\parallel}(z)}\left[\frac{1}{\rho}
\frac{\partial}{\partial \rho} \left(\rho \frac{\partial}{\partial
\rho}\right) + \frac{1}{\rho^2} \frac{\partial^2}{\partial
\theta^2} \right] - \frac{\hbar^2}{2}\frac{\partial}{\partial
z}\frac{1}{m^{\perp}(z)}\frac{\partial}{\partial
z}
\quad\nonumber
\end{eqnarray}
\begin{eqnarray}
 -\frac{i}{2}\hbar \omega_{c}\frac{\partial}{\partial \theta}+\frac{1}{8} m^{\parallel}\omega_{c}^2\rho^2 +
V(\rho,z)\Big\}\Psi(\rho,z)=E\Psi(\rho,z)
 \label{eq2.1},
\end{eqnarray}
where $\omega_c = eB/m^{\parallel}$ is the cyclotron angular
frequency, and $m^\perp$ ($m^\parallel$) is the longitudinal
(in-plane) mass, which depends on $z$, since there is a
heterostructure along this axis. For a QWR with a $R$ radius, the
potential function is defined as $V(\rho,z)=V^{het}(z)$ for $\rho
\leq R$, and $V(\rho,z)=\infty$ otherwise, where $V^{het}(z)$ is
the heterostructure quantum well potential. For numerical
examples, we have used the parameters for GaP/GaAs and InP/InAs
heterostructures, obtained in Ref. 15. For a better description of
the interfacial layer between materials, the existence of graded
interfaces is taken into account, considering a
XP$_\chi$As$_{1-\chi}$ (X = Ga or In) alloy at this region and
assuming that the P composition $\chi$ varies linearly with $z$
from 1 (XP) to 0 (XAs), and that $m^\perp (z) =
m_{XP}^\perp\chi(z) + m_{XAs}^\perp[1-\chi(z)]$ and $V^{het}(z) =
Q_e[\varepsilon_1 \chi(z) + \varepsilon_2 \chi ^2(z)]$, where
$\varepsilon_1$ and $\varepsilon_2$ are interpolation parameters
and $Q_{e}$ is the conduction band offset. \cite{HebertLi} This
approach is similar to the model of Oliveira \emph{et al.},
\cite{Oliveira} but now the in-plane mass is also assumed to
depend on the materials composition $m^\parallel (z) =
m_{XP}^\parallel\chi(z) + m_{XAs}^\parallel[1-\chi(z)]$. It is
important to point out that this one band approach within the
effective mass approximation for heterostructured QWR is good for
studying only conduction band states, while the description of
valence bands states must be made by a $k \cdot p$ multi-band
model. \cite{Voon1}

To solve Eq. (\ref{eq2.1}), we start from a separation of
variables. The solution in $\theta$ is chosen as
$(1/\sqrt{2\pi})e^{il\theta}$, where $l = 0,\pm 1,\pm 2,...$ is
the angular momentum. This leads to
\begin{eqnarray}
\left[-\frac{\hbar^2}{2 m^{\parallel}(z)}\frac{1}{\rho}
\frac{\partial}{\partial \rho} \left(\rho \frac{\partial}{\partial
\rho}\right) + \frac{\hbar^2 l^2}{2 m^\parallel(z)\rho^2} +
\frac{l}{2}\hbar \omega_{c} \quad\nonumber \right.
\end{eqnarray}
\begin{eqnarray}
\left. + \frac{1}{8} m^{\parallel}
\omega_{c}^2 \rho^2\right]
R_{n,l}(\rho)=E_{n,l}^{(\rho)}R_{n,l}(\rho)
 \label{eq3.1}
\end{eqnarray}
for the radial confinement. If $\rho$ is transformed as
$\xi=\rho^2/2a_c$, where $a_c=(\hbar/eB)^{1/2}$ is the cyclotron
radius, Eq. (\ref{eq3.1}) is rewritten as
\begin{equation}
\xi\frac{d^2R(\xi)}{d\xi^2} + \frac{dR(\xi)}{d\xi} -
\left(\frac{l}{2} + \frac{l^2}{4\xi} + \frac{\xi}{4} -
\frac{E}{\hbar\omega_c}\right)R(\xi)=0.
 \label{eq8.1}
\end{equation}

It is reasonable to try a solution of the form
$R(\xi)=\xi^{|l|/2}exp[-\xi/2]F(\xi)$, where the polynomial and
asymptotic behaviors of $R(\xi)$ are explicit. With this solution,
Eq. (\ref{eq8.1}) becomes
\begin{equation}
\xi\frac{d^2F}{d\xi^2} + [(|l|+1) - \xi]\frac{dF}{d\xi}
- \left(\frac{l}{2} + \frac{|l|}{2} + \frac{1}{2} -
\frac{E}{\hbar\omega_c}\right)F=0.
 \label{eq9.1}
\end{equation}

This equation is easily identified as a Confluent Hypergeometric
Equation, namely $xy^{,,} + (c - x)y^,  - ay=0$, which is solved
in terms of Kummer functions. \cite{Spiros} The eigenfunctions are
then found as
\begin{eqnarray}
R_{n,l}(\rho_i)=N\xi^{|l|/2}exp[-\xi/2]F(-x_{n,|l|},|l|+1,\xi),
\label{eq.5.1}
\end{eqnarray}
where $N$ is the normalizing parameter and $F(\alpha,\beta,\xi)$
is the Kummer function of the first kind, which remains finite at
$\xi=0$. From the boundary condition, since $\Psi(R,z_i)=0$, one
has $F(-x_{n,|l|},|l|+1,\xi_R)=0$, hence $x_{n,|l|}$ must be the
$n^{th}$ zero of $F(-x_{n,|l|},|l|+1,R^2/2a_c)$. The radial
confinement energies are given by
\begin{eqnarray}
E_{n,l}^{(\rho)}=\hbar\omega_c\left(x_{n,|l|}+\frac{l}{2}+\frac{|l|}{2}+\frac{1}{2}\right),
\label{eq6.1}
\end{eqnarray}
which clearly depends on $z$, since $\omega_c$ depends on
$m^{\parallel}(z)$. Then, this energy must be added as a potential
in the remaining equation for this coordinate, yielding
\begin{equation}
\left[-\frac{\hbar^2}{2}\frac{\partial}{\partial
z}\left(\frac{1}{m^{\perp}(z)}\frac{\partial}{\partial z}\right) +
V_{eff}(z)\right]Z_m(z)=E_{n,l,m}Z_m(z) , \label{eq4.1}
\end{equation}
where $V_{eff}(z) = V^{het}(z) + E_{n,l}^{(\rho)}(z)$, for the
heterostructure longitudinal confinement, which is solved by a
finite differences scheme.

It is straightforward to show that Eq. (\ref{eq3.1}) becomes the
Bessel equation when $B \rightarrow 0$, giving the Bessel
functions of the first kind as eigenfunctions, \emph{i. e.}
$R_{n,l}(\rho)=J_l(x_{n,l}\rho/R)$, and then the energies are
given by $E_{n,l}^{(\rho)}(z)=\hbar^2x_{n,l}^2/2m^{\parallel}(z)$.
Indeed, the former and the latter functions agree with Eq.
(\ref{eq.5.1}) and Eq. (\ref{eq6.1}), respectively, when the limit
of small $B$ is taken. \cite{Spiros}

\section{Results and Discussions}

We have calculated the electron confinement energies for
cylindrical QWR with longitudinal heterostructures with graded
interfaces, under applied magnetic fields. The material parameters
are considered as $m_{GaAs} = 0.063 m_0$ ($m_{InAs} = 0.027 m_0$)
and $m_{GaP} = 0.33 m_0$ ($m_{InP} = 0.077 m_0$) for electron
effective masses, $\varepsilon_1 = -1.473$ eV and $\varepsilon_2 =
0.146$ eV ($\varepsilon_1 = -1.083$ eV and $\varepsilon_2 = 0.091$
eV) for the interpolation parameters and the conduction band
offset is assumed as $Q_e =$ 0.5 ($Q_e =$ 0.68), for GaP/GaAs
(InP/InAs) heterostructures. \cite{HebertLi}

\begin{figure}[!bpht]
\centerline{\includegraphics[width=.9\linewidth]{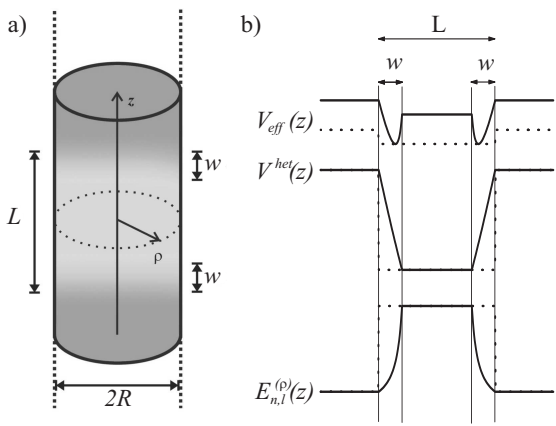}}
\caption{(a) Sketch of a cylindrical QWR with wire radius
\emph{R}, consisting of a single quantum well of width \emph{L}
and considering an interface thickness \emph{w}. (b) Potential
profiles along $z$ for abrupt (dotted) and smooth (solid)
interfaces. The vertical thin lines denote the limits of the
interfacial regions.} \label{fig:6}
\end{figure}

Figure \ref{fig:6}(a) illustrates a representative sketch of our
model of heterostructured QWR. The darker regions represent the
material of the barrier, whereas the lighter one represents the
material that compounds the well. There is a smooth change from
one region to the other, which represents the existence of graded
interfaces. In Fig. \ref{fig:6}(b), a qualitative scheme shows the
potentials $V^{het}(z)$, due to the bands mismatch between the
heterostructure materials, and $E_{n,l}^{(\rho)}(z)$, due to the
lateral confinement, both as functions of $z$. Looking at this
scheme, it seems that $V^{het}(z)$ exhibits linear behavior at the
interfaces but, actually, it is parabolic in $z$ at this region,
as described by the quadratic expression given in Sec. II. In
fact, with this expression, it can be easily shown that $z$ at the
interfacial region is far from the vertex of the parabola
describing the potential at the interfaces and, consequently, the
quadratic curve in this region looks linear. Since the effective
masses at InAs and GaAs are lighter than the ones at InP and GaP,
respectively, the lateral confinement energy $E_{n,l}^{(\rho)}$ is
higher at GaAs (InAs) than at GaP (InP), and hence
$E_{n,l}^{(\rho)}(z)$ presents a barrier like profile, because
this energy depends on the inverse of the effective mass. Besides,
a reduction on the wire radius enhances this barrier like
potential, since $E_{n,l}^{(\rho)}$ also depends on the inverse of
the squared radius. The effective confinement potential
$V_{eff}(z) = V^{het}(z) + E_{n,l}^{(\rho)}(z)$ is also
illustrated in Fig. \ref{fig:6}(b), where one observes that,
depending on the wire parameters, it can exhibit a peculiar form,
where cusps appear at the interfacial regions. The effective
potential $V_{eff}(z)$ is adjusted so that the energy referential
is at the GaP (InP) layers.

\begin{figure}[h!]
\centerline{\includegraphics[width=.9\linewidth]{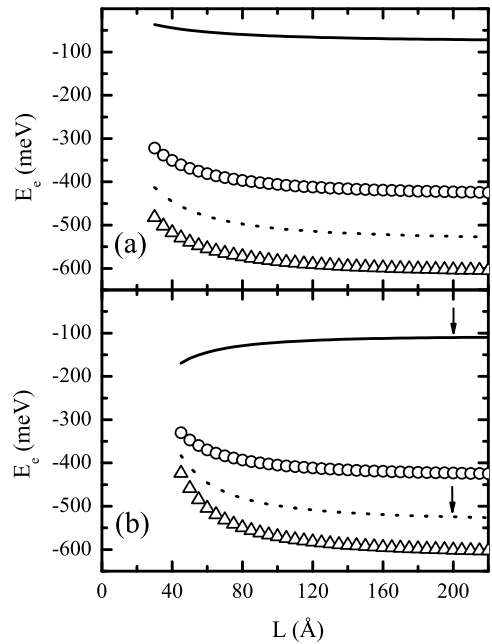}}
\caption{Electron confinement energies in a cylindrical quantum
wire with a longitudinal GaP/GaAs heterostructure, with interface
thickness (a) 0 \AA\, and (b) 20 \AA\,, as a function of the well
width $L$, for $l$ = 0 (symbols) and $l$ = 1 (curves). Two values
of the wire radius $R$ were considered: 35 \AA\,($\circ$, solid)
and 75 \AA\,($\triangle$, dotted). The electron wavefunctions for
the points indicated by arrows ($L$ = 200 \AA\,) are plotted in
Fig. 4(a).} \label{fig:4}
\end{figure}

\begin{figure}[h!]
\centerline{\includegraphics[width=0.9\linewidth]{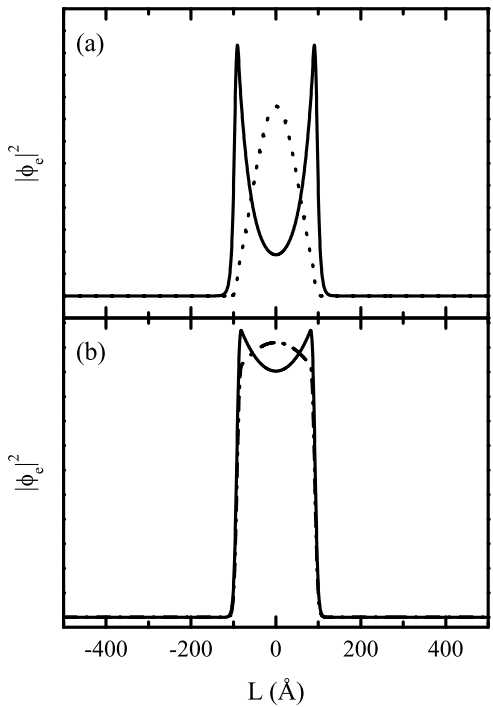}}
\caption{Electron wave functions for $l$ = 1 states in a
cylindrical GaP/GaAs QWR with graded interfaces of $w$ = 20 \AA\,
thickness and well width $L$ = 200 \AA\,, as a function of $z$,
considering (a) wire radius $R$ = 35 \AA\, (solid) and $R$ = 75
\AA\, (dotted) in the absence of magnetic fields; and (b) magnetic
fields $B$ = 0 T (dashed-dotted) and $B$ = 30 T (solid) for a $R$
= 42 \AA\, wire radius.} \label{fig:14}
\end{figure}

Figure \ref{fig:4} shows the confinement energies $E_{n,l,m}$ of
electrons in the absence of magnetic fields, for $E_{1,0,1}$
(symbols) and $E_{1,1,1}$ (curves) states in a GaP/GaAs QWR, as a
function of the well width $L$, formed by the longitudinal
heterostructure. We have considered interface thicknesses of (a)
$w$ = 0 \AA\, and (b) 20 \AA\, between materials, and two values
for the wire radius $R$: 35 \AA\,($\circ$, solid) and 75
\AA\,($\triangle$, dotted), which are a small and a large value of
wire radius, in order to observe the different behaviors due to
stronger or weaker bidimensional confinements. The electron
wavefunctions for $l = 1$ states in a $L = 200$ \AA\, non-abrupt
QWR with $R = 35$ \AA\, and $R = 75$ \AA\, (points marked by
arrows) are plotted in Fig \ref{fig:14} (a) for further analysis.
For abrupt interfaces, all energies decrease as the well width $L$
is enlarged, as usually observed in quantum wells. However, when
graded interfaces are taken into account in a $R = 35$ \AA\, wire,
the energy of the electron $l = 1$ state increases with the well
width, for small values of $L$. This indicates that this state is
now confined at the interfacial regions, so that an enlargement on
the well width $L$ further separates the interfaces and increases
the confinement energy of the $n = 1$ state, analogous to the case
of confinement in double quantum wells. \cite{Bastard} On the
other hand, $l = 0$ states for $R = 35$ \AA\, and the states for
$R = 75$ \AA\, keep the same qualitative behavior for abrupt and
$w = 20$ \AA\, interfaces, \emph{i. e.} their energies decrease as
increasing $L$, but quantitatively, the presence of graded
interfaces still plays an important role, giving a significative
blueshift on these energies, specially for small $L$ and large
$R$. As a numerical example, for $L = 45$ \AA\,, the energy
blueshift due to the graded interfaces, given by $\Delta E_{e} =
E_e(w = 20) - E_e(w = 0)$, is about $\Delta E_{e}$ $\simeq$ 30 meV
(105 meV) for $l = 0$ states in a $R = 35$ \AA\, (75 \AA\,) QWR.
This blueshift in confinement energies for non-abrupt
heterostructures has also been predicted in quantum wells,
\emph{core-shell} QWR and quantum dots with graded interfaces.
\cite{JAPChaves, PRBChaves, PECaetano} In fact, for large wire
radius, $E_{n,l}^{(\rho)}(z)$ is small, thus the presence of
graded interfaces only reduces the confinement region in
$V_i^{het}(z)$, enhancing the energy levels. Yet, for small $R$,
the presence of an interfacial region creates cusps that can
confine carriers (see Fig. \ref{fig:6} (b), solid), which are
responsible for the reduction on the energy blueshift due to
interfaces in this case, and can even lead to a redshift, as
observed for $l = 1$ when $R = 35$ \AA\,.
\begin{figure}[h!]
\centerline{\includegraphics[width=0.9\linewidth]{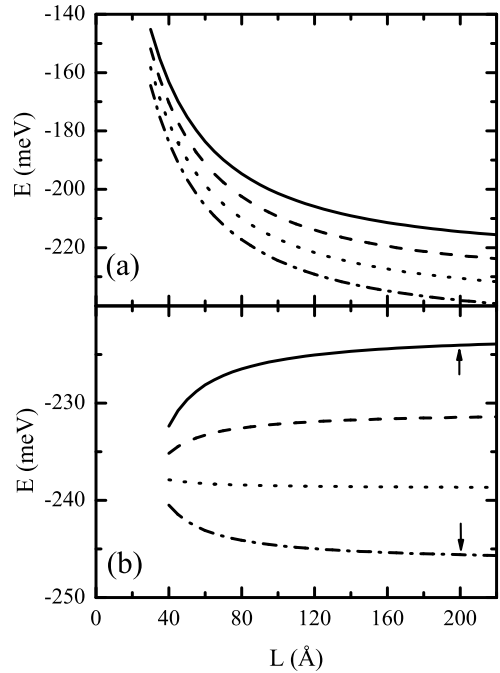}}
\caption{Electron confinement energies as functions of the well
width $L$ for $l$ = 1 states in a cylindrical GaP/GaAs QWR with
wire radius $R$ = 42 \AA\,, considering (a) abrupt and (b) $w$ =
20 \AA\, interfaces, under applied magnetic fields $B$ = 0 T
(dashed-dotted), 10 T (dotted), 20 T(dashed) and 30 T (solid). The
electron wavefunctions for the points indicated by arrows ($L$ =
200 \AA\,) are plotted in Fig. 4(b).} \label{fig:5}
\end{figure}

As was just discussed, the interfacial confinement for electron
lower energy states, as well as the type-I (well) to type-II
(barrier) transition in effective potential predicted in previous
works \cite{Voon1}, are found for small values of $R$, which can
be troublesome for experimental verification of these features of
QWR. Nevertheless, one could try to find another way to induce
these variations on the carrier´s localization: The main effect
produced by reducing the QWR radius is that the carrier´s wave
functions are squeezed towards the wire axis. If a magnetic field
is applied parallel to the wire axis, the same effect can be
obtained, hence, type-I to type-II transitions and interfacial
confinements, which were found for wires with small radii, are
expected to be found for high magnetic fields as well.

Figure \ref{fig:5} is then devoted to the study of the influence
of an applied magnetic field, parallel to the wire axis, on the
confinement energies of electrons. The energies of $l = 1$ states
in a GaP/GaAs QWR with $R = 42$ \AA\, are shown as functions of
the well width $L$, considering (a) $w = 0$ \AA\, and (b) $w = 20$
\AA\,. Four values of magnetic field are considered: $B$ = 0 T
(dashed-dotted), 10 T (dotted), 20 T (dashed) and 30 T (solid).
The $l = 0$ states are not shown, because the dependence of such
states on the magnetic field is negligible; $l = 1$ states are
much more affected by increasing this field, since in the
Hamiltonian of Eq. (\ref{eq3.1}) there is an additional term
involving the cyclotron frequency $\omega_c$ and the angular
momentum $l$. The electron wavefunctions of $l = 1$ states in the
non-abrupt case with $L = 200$ \AA\,, for $B = 0$ T and $B = 30$ T
(points marked by arrows), are plotted in Fig \ref{fig:14} (b) for
further analysis. For abrupt interfaces (Fig. \ref{fig:5} (a)),
the presence of a magnetic field enhances the confinement
energies, but gives no appreciable change in qualitative behavior
of $E_e \times L$ curves. However, considering graded interfaces
(Fig. \ref{fig:5} (b)), these curves are qualitatively different:
the energy behavior for $B = 20$ T (dashed) and $B = 30$ T (solid)
is crescent as the well width $L$ increases, whereas the opposite
behavior is observed for $B = 0$ T (dashed-dotted) and $B = 10$ T
(dotted). This can be understood as an interfacial confinement of
these states, not induced by reducing the wire radius, as in Fig.
\ref{fig:4} (b), but due to the presence of a magnetic field
parallel to the wire axis. Numerical results obtained from our
model demonstrate that, keeping the same interface thickness $w =
20$ \AA\,, but considering a slightly smaller radius, $R = 40$
\AA\,, the electron would be already confined at the interface, so
that there is no need for applying a magnetic field to observe
this kind of confinement. However, with $R = 40$ \AA\, and a
smaller interface thickness $w = 15$ \AA\,, the electron is
confined at the GaAs layer and, in this case, interfacial
confinements induced by magnetic fields were also found, but only
for $B \gtrsim 30$ T, which is higher than the values for $R = 42$
\AA\, and $w = 20$ \AA\, shown in Fig \ref{fig:5} (b), where a $B
= 20$ T magnetic field was enough for inducing an interfacial
confinement. Our results also predict that interfacial confinement
of $l = 1$ electrons due to magnetic fields can be found in
InP/InAs QWR too, with a slightly larger radius $R = 48$ \AA\,
(not shown in this work). This can be an interesting feature of
these systems since, once the QWR is grown its radius is fixed and
an external parameter, namely the magnetic field, can just be
tuned to obtain an electron confinement at the well or at the
interfaces.

\begin{figure}[b!]
\centerline{\includegraphics[width=0.9\linewidth]{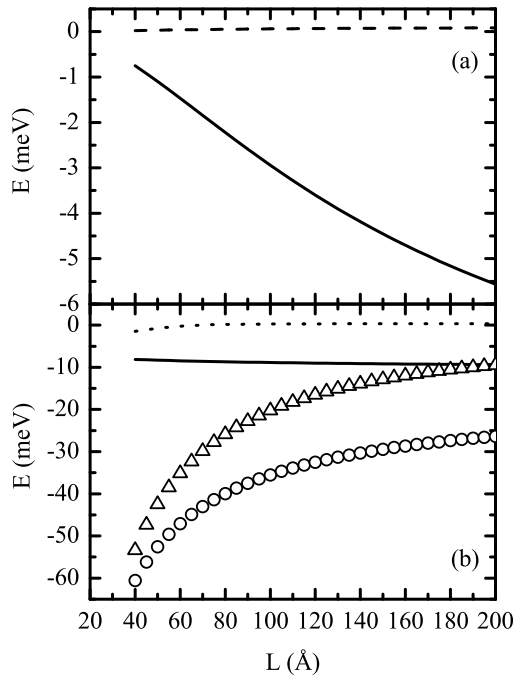}}
\caption{(a) Electron confinement energies as functions of the
well width $L$ for $l$ = 1 states in a cylindrical InP/InAs QWR
with wire radius $R$ = 45 \AA\, and abrupt interfaces, under
applied magnetic fields $B$ = 0 T (solid) and 10 T(dashed). (b)
The same of (a), but considering graded interfaces of $w$ = 5
\AA\, (curves) and $w$ = 20 \AA\, (symbols), and magnetic fields
$B$ = 0 T (solid, $\circ$) and 20 T(dotted, $\triangle$).}
\label{fig:15}
\end{figure}

The interfacial confinement of states in GaP/GaAs QWR can be
verified by analyzing the electron wave functions as a function of
$z$ in such systems, which is illustrated in Fig. \ref{fig:14},
for $E_{1,1,1}$ states. In Fig. \ref{fig:14}(a), two values of
wire radius are considered, in the absence of magnetic fields,
while in (b) the wire radius is kept and two values of magnetic
field are considered. It can be observed in Fig. \ref{fig:14}(a)
that a confinement of $l$ = 1 states at the interfaces is induced
when the wire radius is reduced from $R$ = 75 \AA\, (dotted) to
$R$ = 35 \AA\, (solid). When a magnetic field is applied, in Fig.
\ref{fig:14}(b), the wave function of such states in a $R$ = 42
\AA\, QWR is also altered, and for $B$ = 30 T (solid), two peaks
at the interfacial regions can be seen. Thus, these results
confirm that, in the presence of graded interfaces, confined
states at the interfacial regions are observed by reducing the
wire radius or increasing the magnetic field intensity.

In Fig. \ref{fig:15}, we present the electron confinement energies
as a function of the well width $L$, for $l = 1$ states in
InP/InAs QWR with $R = 45$ \AA\,. These results demonstrate that
the type-I to type-II transition predicted in previous work on
abrupt wires can also be obtained by increasing the intensity of a
magnetic field parallel to the wire axis: for $w = 0$ \AA\, (Fig.
\ref{fig:15} (a)), the confinement energies for $B = 0$ (solid)
decrease with increasing $L$, but for $B = 10$ T (dashed)
$E_{1,1,1} = 0$ for all $L$, which shows that these states are not
confined in $z$, despite the presence of a heterostructure. This
indicates that the radial confinement energy $E_{n,l}^{(\rho)}(z)$
for $B = 10$ T, which is a barrier like potential, is high enough
to suppress the contribution of the heterostructure potential in
$V_{eff}(z)$ for this system, yielding a type-II effective
potential. Hence, with a InP/InAs QWR with abrupt interfaces, one
could control the electron band offset or even change from a
confinement to a scattering potential for $l \neq 0$ states, just
by setting the external magnetic field. However, real QWR are
shown to present graded interfaces, and taking them into account
in Fig. \ref{fig:15} (b), the results show that if one considers
an interface thickness $w = 20$ \AA\, (symbols) \emph{e. g.}, the
magnetic field does not change the qualitative behavior of $l = 1$
electrons, which are confined at the interfaces for both $B = 0$ T
($\circ$) and 20 T ($\triangle$). For a small interface thickness
$w = 5$ \AA\, (curves), in the absence of magnetic fields (solid),
the electron $l = 1$ state is confined at the InAs (well) layer,
and it depends weakly on the well width, due to the new form of
$V_{eff}(z)$: for $L$ varying from 40 \AA\, to 200 \AA\,
considering $w = 0$ \AA\,, $E_{1,1,1}$ decreases in $\sim$ 5 meV,
whereas for $w = 5$ \AA\,, it varies only $\sim$ 1 meV. For a $B =
20$ T magnetic field, considering $w = 5$ \AA\, (dotted), an weak
interfacial confinement of $l = 1$ electrons is still observed for
$L$ lower than $\sim 80$ \AA\,, while for greater values of $L$,
the barrier on the effective potential is large enough to avoid
confinement at the interfaces, leading to $E_{1,1,1} = 0$ with an
electron localization at the InP layers, just like in a type-II
system. Thus, for a perfect type-I to type-II transition induced
by a magnetic field in InP/InAs non-abrupt QWR, one would need a
$R = 45$ \AA\, wire with a large InAs layer and small interfaces
($<$ 5 \AA\,), \emph{i. e.} with a very high quality of
heterostructure modulation, in order to avoid interfacial
confinements.
\begin{figure}[h!]
\centerline{\includegraphics[width=0.9\linewidth]{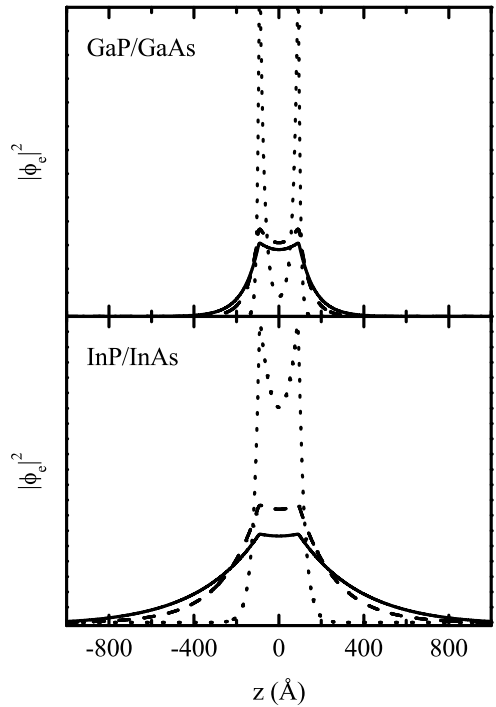}}
\caption{Electron wave functions for $l$ = 1 states in cylindrical
GaP/GaAs and InP/InAs QWR with graded interfaces of $w$ = 20 \AA\,
thickness and well width $L$ = 200 \AA\,, as a function of $z$,
considering three different approximations for the effective mass
variation on the interfacial region: direct (dotted) and
reciprocal (solid) mass variations considering $\chi(z)$ as an
error function, and a variation on the reciprocal mass with a
linear function for $\chi(z)$ (dashed). The wire radius is $R$ =
33 \AA\, ($R$ = 45 \AA\,) for GaP/GaAs (InP/InAs).} \label{fig:16}
\end{figure}

The presence of magnetic fields has been shown to induce
interfacial confinement for $l = 1$ states, because this field
enhances the radial confinement energy $E_{n,l}^{(\rho)}$ and then
supports this change of localization. However, for $l = -1$
states, the magnetic field reduces $E_{n,l}^{(\rho)}$, hence, it
would never induce a change from well to interfacial localization
of such states. Conversely, there are some cases where the $l =
-1$ electron state is already confined at the interfaces, for
instance, in a GaP/GaAs QWR with $R = 39 $ \AA\, and $w = 15$
\AA\,, and a magnetic field $B = 30$ T can induce a transition
from interfacial to well localization. Such a transition can also
be obtained with $R = 40 $ \AA\, and $w = 20$, but applying a
higher magnetic field intensity, $B = 35$ T.

Since the mass variation through the interfacial regions is of
major importance for the effective potential and, consequently,
for the carriers confinement, it is worthwhile to discuss on the
possible types of function which can be used to describe such a
variation. In present work, as one can verify in Sec. II, the
interfacial region is assumed to be a XP$_{\chi}$As$_{1-\chi}$ (X
= Ga or In) alloy, with a P composition $\chi(z)$ varying linearly
along $z$ at the interfaces. The effective masses are then assumed
to depend linearly on $\chi(z)$. However, other dependencies on
$\chi(z)$, for instance, a linear dependence for the reciprocal
effective mass, \emph{i. e.}, $1/m(z) = (1/m_{XP})\chi(z) +
(1/m_{XAs})[1-\chi(z)]$ could be considered. The following
procedure shows straightforwardly that even for this kind of
variation, one can still tailor the system in order to find
confining potentials at the interfaces: Suppose one has a linear
function $\chi(z) = z/w$, describing a composition variation at an
interface lying within $0 \leqslant z \leqslant w$. The effective
mass and the heterostructure potential are then given by $1/m(z) =
A(z/w)+(1/m_{XAs})$ and $V^{het}(z) = Q_e[\varepsilon_1 (z/w) +
\varepsilon_2 (z/w)^2]$, respectively, with $A = 1/m_{XP} -
1/m_{XAs}$. The effective potential in this region is then
\begin{eqnarray}
V_{eff}(z) = V^{het}(z) + E_{n,l}^{(\rho)}(z) = \nonumber \\
a\frac{z}{w} +
b\left(\frac{z}{w}\right)^2 +
C\left(A\frac{z}{w}+\frac{1}{m_{XAs}}\right),
\end{eqnarray}
where $a = Q_e\varepsilon_1$, $b = Q_e\varepsilon_2$ and $C$ is
obtained from Eq. (\ref{eq6.1}) as $ C = \hbar
eB[x_{n,|l|}+(l/2)+(|l|/2)]$. A minimum value of $V_{eff}(z)$
exists within $0 \leqslant z \leqslant w$ if $dV_{eff}/dz = 0$
somewhere in this interval. After some manipulation of the
equations, one eventually finds that a minimum value of
$V_{eff}(z)$ can still occur at the interface for this kind of
mass variation, provided $|(CA+a)/(2b)| < 1$. The parameters $A$,
$a$ and $b$ are constants of the material, while $C$ is a function
of the wire radius $R$ that can be tuned so that this inequality
is satisfied. This inequality would be satisfied by no value of
$C$ only if $b =$ 0, \emph{i. e.}, $\varepsilon_2 =$ 0 and this
means that the energy gap of the alloy depends linearly on the
composition $\chi$, which is not true for most of the materials in
the literature. \cite{HebertLi} Figure \ref{fig:16} shows the
electron wavefunctions along $z$ for $l = 1$ states in GaP/GaAs
and InP/InAs QWR with well width $L = 200$ \AA\, and interfaces
thickness $w = 20$ \AA\,. Dashed line is obtained by considering
$\chi(z)$ as a linear function of $z$ and a variation on the
reciprocal effective mass $1/m$, whereas the other curves depict
the results considering $\chi(z)$ as an error function
\cite{Chaves} of $z$, with variations on the direct $m$ (dotted)
and reciprocal $1/m$ (solid) effective masses. Electron
confinement at interfaces is observed in all the curves shown,
although for the reciprocal mass cases the electron is just weakly
bound in this region and is more dependent on the wire parameters:
when the reciprocal mass varies in GaP/GaAs (InP/InAs), the
interfacial confinement was found only for $R \sim$ 33 \AA\,
($\sim$ 45 \AA\,), whereas for a variation in the direct mass,
used in our model to obtain the previous results of this paper, an
interfacial localization is observed for radii varying within a
range from $R \sim$ 27 \AA\, to $\sim$ 40 \AA\, ($R \sim$ 42 \AA\,
to $\sim$ 46 \AA\,).

The theoretical model we suggest in this work is good for the
description of electron states in conduction band, but it fails
when studying valence band states because in cases where the wire
diameter $2R$ and the heterostructure width $L$ have almost the
same dimensions, the lowest valence state is a combination of
heavy-hole (HH) and light-hole (LH) states. Hence, to correctly
solve the QWR problem for holes when $2R \simeq L$, one must use a
$4 \times 4$ Hamiltonian, taking into account HH and LH states.
Even so, based on the fact that the changes in localization are
strongly dependent on the differences between effective masses, it
can be inferred that the critical radii for type-I to type-II
transitions, as well as those for interfacial confinements, are
not expected to be the same for electrons and holes, since the
effective masses of these carriers are completely different in
each material. Indeed, in a previous work, L. C. Lew Yan Voon
\textit{et al.} \cite{Voon1} have used a four-band $k \cdot p$
based theory to predict type-I to type-II transitions also for
valence band states in abrupt InGaAs/InP and GaAs/AlAs QWR, and
they found that critical radii are different for electrons and
holes. Furthermore, considering a GaAs/GaP QWR, changes in valence
band states localization would hardly be obtained, due to the
small difference between the hole effective masses in GaAs and
GaP. However, changes in the electrons localization are found in
this system (see Fig. \ref{fig:4}(b)). These electron-hole
separations might decrease the overlap between their ground state
wave functions, which would reduce the probability of an interband
transition for such states, a feature that is also commonly found
in type-II systems. \cite{Voon2} Thus, a low probability of
interband transitions would be observed in cases where the hole
remains at the well layer and the electron is confined at the
interfaces or vice-versa.

\section{Conclusions}

The confinement energies of electrons in cylindrical nanowires
with longitudinal GaP/GaAs and InP/InAs heterostructures, under an
applied magnetic field parallel to the wire axis, were calculated
for several values of wire radius, well width, interfaces
thickness and magnetic field intensity. The difference between
effective masses of the heterostructure materials plays an
important role in the carrier confinement along the wire axis; it
is responsible for creating a new $z$-dependent potential, since
the lateral confinement energy depends on such masses. It was
demonstrated that, for abrupt interfaces, reducing the wire radius
leads to a weaker effective confinement potential along the wire
axis, due to an enhancement of the radial confinement energy. The
energies of $l = 1$ states are lower than the ones for $l = 0$
states, which is due to the fact that the former presents higher
radial confinement energy, and consequently the effective
confinement potential along the wire axis is reduced. The
existence of smooth interfaces between materials drastically
changes the effective potential profile and can yield to a carrier
confinement at the interfacial region for thin wires with $w \neq
0$, instead of the free carrier behavior or the type-I to type-II
transitions predicted in previous works on abrupt QWR.
\cite{Voon1, Voon3} Several kinds of functions were considered for
the description of the effective mass variation at the interfaces,
and our results demonstrate that an interfacial confinement of
carriers can be found for all cases shown. The confinement
energies are slightly affected by the presence of magnetic fields,
particularly for $l = 0$ states, but in some special cases,
increasing the magnetic field intensity also induces a type-II
potential, in the abrupt case, or a carrier confinement at the
interfaces, when $w \neq 0$. This can be a useful feature of these
systems for device applications, since it shows that the carrier
behavior can be controlled by an external field. Our results also
suggest that in some cases of heterostructured QWR, interband
transitions may be strongly suppressed, due to spatial separation
of electrons and holes states. Finally, the features shown in
present work allow us to infer that a high control of composition
and size parameters is essential to the design and fabrication of
heterostructured quantum wire based devices.

\acknowledgements Discussions with F. M. Peeters are gratefully acknowledged. This work has received financial support from
the Brazilian National Research Council (CNPq), under contract
NanoBioEstruturas 555183/2005-0, Funda\c{c}\~ao Cearense de Apoio
ao Desenvolvimento Cient\'ifico e Tecnol\'ogico (Funcap) and
Pronex/CNPq/Funcap.


\begin{references}

\bibitem{review} M. Law, J. Goldberger and P. Yang, Annu. Rev. Mater. Res.
\textbf{34}, 83 (2004).
%
\bibitem{Deppert} M. T. Bj\"ork, B. J. Ohlsson, T. Sass, A. I. Persson, C.
Thelander, M. H. Magnusson, K. Deppert, L. R. Wallenberg and L.
Samuelson, Nano Lett. \textbf{2}, 87 (2002)
%
\bibitem{Solanki} R. Solanki, J. Huo, J. L. Freeouf and B. Miner, Appl. Phys. Lett. \textbf{81}, 3864
(2002).
%
\bibitem{Gudiksen} M. S. Gudiksen, L.  J. Lauhon, J. Wang, D. C. Smith and C. M.
Lieber, Nature \textbf{415},  617 (2002).
%
\bibitem{Chaves} C. L. N. Oliveira, A. Chaves, E. W. S. Caetano, M. H.
Degani and J. A. K. Freire, Microelectr. J. \textbf{36}, 1049
(2005).
%
\bibitem{Li} D. Li, Y. Wu, R. Fan, P. Yang and A. Majumdar, Appl.
Phys. Lett. \textbf{83}, 3186 (2003).
%
\bibitem{Dresselhaus} Y.-M. Lin and M. S. Dresselhaus, Phys. Rev.
B \textbf{68} 075304 (2003).
%
\bibitem{Wu} Y. Wu, R. Fan and P. Yang, Nano Lett. \textbf{2}, 83 (2002).
%
\bibitem{Fuhrer} A. Fuhrer, L. E. Fr\"oberg, J. N. Pedersen, M. W. Larsson, A.
Wacker, M.-E. Pistol and L. Samuelson, Nano Lett. \textbf{7}, 243
(2007).
%
\bibitem{Voon2} L. C. L. Y. Voon and M. Willatzen, J.
Appl. Phys. \textbf{93}, 9997 (2003).
%
\bibitem{Weiser} H. Gr\"uning, P.J. Klar, W. Heimbrodt, S. Nau, B. Kunert, K.
Volz, W. Stolz and G. Weiser, Physica E \textbf{21}, 666 (2004).
%
\bibitem{Voon3} M. Willatzen, R. V. N. Melnik, C. Galeriu and L. C. L. Y. Voon, Math. Comput.
Simulat. \textbf{65}, 385 (2004).
%
\bibitem{Voon1} L. C. L. Y. Voon, B. Lassen, R. Melnik and M.
Willatzen, J. Appl. Phys. \textbf{96}, 4660 (2004).
%
\bibitem{king} E. C. Ferreira, J. A. P. da Costa and J. A. K.
Freire, Physica E \textbf{17}, 222 (2003).
%
\bibitem{HebertLi} E. H. Li, Physica E \textbf{5}, 215 (2000).
%
\bibitem{Oliveira} {C. L. N. Oliveira, J. A. K. Freire, V. N. Freire and G. A. Farias},
Appl. Surf. Sci.  \textbf{234}, 38 (2004).
%
\bibitem{Spiros} S. V. Branis, G. Li and K. K. Bajaj, Phys. Rev. B \textbf{47}, 1316 (1993).
%
\bibitem{Bastard} R. Ferreira and G. Bastard, Rep. Prog. Phys. \textbf{60}, 345 (1997).
%
\bibitem{JAPChaves} A. Chaves, J. Costa e Silva, J. A. K. Freire and G. A. Farias,
J. Appl. Phys. \textbf{101}, 113703 (2007).
%
\bibitem{PRBChaves} J. Costa e Silva, A. Chaves, J. A. K. Freire, V. N. Freire and G. A. Farias,
Phys. Rev. B \textbf{74}, 085317 (2006).
%
\bibitem{PECaetano} E. W. S. Caetano, M. V. Mesquita, V. N. Freire and G. A. Farias,
Physica E \textbf{17}, 22 (2003).


\end{references}
\end{document}